\documentclass[aps,prb,amsmath,amssymb,longbibliography,lengthcheck,showpacs,superscriptaddress,floatfix]{revtex4-1}
\usepackage{graphicx}
\usepackage{color}
\begin{document}
\title{Sound damping in glasses: Interplay between anharmonicities and elastic heterogeneities}
\author{Hideyuki Mizuno}
\email{hideyuki.mizuno@phys.c.u-tokyo.ac.jp}
\affiliation{Graduate School of Arts and Sciences, The University of Tokyo, Tokyo 153-8902, Japan}
\author{Giancarlo Ruocco}
\affiliation{Dipartimento di Fisica, Sapienza Universit\`a di Roma, Piazzale A. Moro 2, I-00185, Rome, Italy}
\affiliation{Center for Life Nano Science, Istituto Italiano di Tecnologia, Viale Regina Elena 291, I-00161, Rome, Italy}
\author{Stefano Mossa}
\email{stefano.mossa@cea.fr}
\affiliation{Univ. Grenoble Alpes, CEA, IRIG-MEM, 38000 Grenoble, France}
\affiliation{Institut Laue-Langevin, BP 156, F-38042 Grenoble Cedex 9, France}
\date{\today}
\begin{abstract}
Some facets of the way sound waves travel through glasses are still unclear. Recent works have shown that in the low-temperature harmonic limit a crucial role in controlling sound damping is played by local elastic heterogeneity. Sound waves propagation has been demonstrated to be strongly affected by inhomogeneous mechanical features of the materials, which add to the anharmonic couplings at finite temperatures. We describe the interplay between these two effects by molecular dynamics simulation of a model glass. In particular, we focus on the transverse components of the vibrational excitations in terms of dynamic structure factors, and characterize the temperature dependence of sound attenuation rates in an extended frequency range. We provide a complete picture of all phenomena, in terms encompassing both theory and experiments.
\end{abstract}
\maketitle

\section{Introduction}
At vanishing temperatures, the harmonic approximation describes vibrations in crystals as a collection of non-interacting quasi-particles with well-defined energy {\em and} momentum, the phonons~\cite{Ashcroft}, providing a systematic reference state for any further description. At non-zero temperatures, {\color{black} couplings due to the anharmonicities of the interaction potential} trigger the insurgence of finite life-times of phonons, which can be described via the Boltzmann transport equation~\cite{McGaughey}. The phenomenology is substantially richer for glasses~\cite{zeller1971thermal,pohl2002low,klinger2010soft}. At low temperatures and small frequencies (wave-numbers), where the continuum limit holds, the phonon-like picture is still helpful. In contrast, when phenomena occurring at length scales comparable to the atomic distance are involved~\footnote{Note that other excitations, including two-level systems (see, among many others,~\cite{leggett2013tunneling}), are in principle relevant. We do not consider these issues here.}, additional concepts are needed~\cite{lowtem,Isaeva_2019,simoncelli2019unified}. Indeed, the presence of structural disorder now imposes that, while the quasi-particles have well-defined energy, their momentum is ill-defined. On increasing temperature, when the strength of anharmonicities grows, the situation becomes even more complex~\cite{mizuno2019anharmonic}. The interplay between disorder and anharmonicity is still a rather unexplored issue.

Numerous experiments have demonstrated anomalous transport of acoustic-like excitations in glasses in the THz-GHz regime, including breakdown of the Debye approximation (sound softening)~\cite{monaco_2009} and Rayleigh-like strong scattering, which determines an apparent life-time $\tau$ ($1/\tau=\Gamma\propto\Omega^4$)~\cite{masciovecchio2006evidence,devos2008hypersound,baldi2010sound}. In~\cite{monaco2009anomalous}, for instance, some of us highlighted by Molecular Dynamics (MD) simulation a crossover from the Rayleigh-like scattering to a disorder-induced broadening, $\Gamma\propto\Omega^2$, at higher frequencies. Remarkably, we found that the crossover frequency for transverse excitations, $\Omega_\text{co}$, is close to the Ioffe-Regel limit $\Omega_\text{IR}$, indicating the sound waves start to lose their plane-wave character at $\Omega_\text{co}$. In addition, in the same frequency regime, non-conventional features, such as the excess vibrational intensity of the Boson peak (BP)~\cite{buchenau_1984,Malinovsky_1991} and vibrational localization~\cite{mazzacurati_1996,Schober_2004}, were observed. Since sound waves in glasses can be described as envelopes of vibrational modes~\cite{taraskin_2000}, these properties are closely related to the anomalous sound waves propagation. In particular, an universal connection of transverse sound waves with the Boson peak has been proposed~\cite{monaco2009anomalous,Shintani_2008}.

We can rationalize these issues in terms of a local {\em elastic heterogeneity}~\cite{Duval_1998}. Recent simulation~\cite{yoshimoto_2004,tsamados_2009,makke_2011} and experimental~\cite{Wagner_2011,Hufnagel_2015} works have demonstrated that glasses exhibit inhomogeneous mechanical response at the nano-scale, i.~e., elastic moduli do not simply assume the hydrodynamic values but rather fluctuate around it, with a finite distribution width. This subtle heterogeneity generates in turn {\em non-affine} deformations~\cite{DiDonna_2005,Maloney_2006}, which add to the applied affine field inducing a significant reduction in elastic moduli~\cite{tanguy_2002,Zaccone_2011}. Following the non-affine deformation, particles turn out to be displaced in a correlated manner, characterized by a typical mesoscopic correlation length~\cite{leonforte_2005}. It is natural to expect that interaction with the non-affine displacement field modifies sound propagation. In~\cite{Mizuno2_2013,Mizuno_2014,Mizuno_2016}, we have provided strong evidences of this direct correlation between sound waves features and the heterogeneous mechanical properties. Also based on these ideas, Schirmacher {\em et al.}~\cite{schirmacher_2006,schirmacher_2007,Marruzzo_2013,Schirmacher_2015} have developed a heterogeneous elasticity theory which reproduces numerous of the above features. 

Comprehensibly, a large part of computational investigation on these issues, has tended to focus on quasi-harmonic (inherent structures) conditions at zero temperature~\cite{Gelin_2016,Bouchbinder_2018,mizuno2018phonon,angelani2018probing,wang2019sound,Moriel_2019}. How these mechanical features entangle with anharmonicities determining the total sound attenuation is therefore an open issue not yet extensively explored~\cite{schirmacher2010sound,Tomaras_2010}. Here, we address the interplay of both anharmonic couplings {\em and} elastic disorder in a standard atomic glass. In particular, we focus on the transverse component of the dynamic structure factor, by simulating extremely large glassy samples, and analyze the attenuation rates in an extended frequency range, at varying temperatures. By disentangling the different interaction channels, we describe in details and in a unified perspective the main scattering mechanisms. 

\section{Methods}
%
\subsection{System description}
We have studied by MD simulations glassy systems formed by $N$ mono-dispersed particles, of mass $m$ and diameter $\sigma$, interacting via the LJ potential, 
\begin{equation}
V(r)=4\epsilon \left[ \left( \frac{\sigma}{r}\right)^{12}-\left(\frac{\sigma}{r}\right)^6 \right],
\end{equation}
where $r=r_{ij}$ the distance between particles $i$ and $j$. $V(r)$ is cut-off and shifted at $r_c=2.5\,\sigma$~\footnote{We are aware that discontinuity of the derivative $V'(r)$ at $r_c$ can, in general, modify some vibrational features. Based on previous work (see, for instance, our~\cite{mizuno2016cutoff}), however, we do not expect relevant modifications in this case.}. We have employed cubic boxes of size $L$, with number density $\hat{\rho}=N/L^3=1.015$. At $\hat{\rho}$, the melting and glass transition temperatures are $T_m \simeq 1.0$ and $T_g \simeq 0.4$, respectively~\cite{robles2003liquid}. In order to access the small wave-number ($q$) region relevant here, we have considered eight values of $N$ ranging from $4000$ to $1000188$, which correspond to values of $L$ in the range $15.80$ to $99.51$. In the following we show data pertaining to all values of $N$ together, directly verifying the absence of any finite size effects. 

Initialization runs were conducted at temperature $T=2.0$ in the normal liquid phase, followed by a fast quench rate $dT/dt\approx 400$ down to $T = 10^{-3}$. Next the systems were heated to $T=10^{-2}$, $10^{-1}$, $2\times 10^{-1}$, still below $T_g$. Following thermalization, we performed the production runs for a ($N$-dependent) total time sufficient to obtain the desired $\omega$-resolution, always well below the smallest calculated line widths. Here we emphasize that there are no aging effects recognized during the productions runs at least in the time history of total energy. We used LAMMPS~\cite{plimpton1995fast} for our runs, and the reader can refer to~\cite{mizuno2019impact} for all additional details.

\subsection{Analysis of sound propagation}
Sound propagation has been investigated in terms of the transverse dynamical structure factors~\cite{monaco2009anomalous,Mizuno_2014} at wave numbers ${q}$ and frequencies $\omega$:
\begin{equation} 
\label{eq:stqomega}
S_T(q,\omega) = \frac{1}{2 \pi N} \left( \frac{q}{\omega} \right)^2 \int dt \left< \mathbf{j}_{T}(q,t)\cdot\mathbf{j}^{\dagger}_{T}(q,0) \right> e^{i\omega t},
\end{equation}
where $\mathbf{j}_{T}(q,t) = \sum_{i=1}^{N} \left\{ \mathbf{v}_i(t) - \left[ \mathbf{v}_i(t) \cdot \widehat{\mathbf{q}} \right] \widehat{\mathbf{q}} \right\} \exp\{i \mathbf{q} \cdot \mathbf{r}_i(t) \}$ is the transverse current vector. Here, $q=\vert \mathbf{q} \vert$, $\hat{\mathbf{q}} = \mathbf{q}/q$, and $\left< \right>$ is the thermodynamic average. Although inelastic experiments with Neutrons and X-Rays probe the longitudinal component of $S_L(q,\omega)$, it has demonstrated that the transverse counterpart follows very similar patterns at higher values of $q$ and $\omega$, making the computations more comfortable.
In the paper we have systematically dumped the subscript ($T$), and have indicated frequencies with $\Omega$.

The spectra were complemented by the vDOS, $g(\omega)$, determined by numerically diagonalizing the Hessian matrix. We have used different system sizes, up to $N=256000$, in order to adequately sample the $g(\omega)$. For our LJ glass, $q_D = (6\pi^2\hat\rho)^{1/3} \simeq 3.92$, $c_D = \left[ \left(c_L^{-3}+2 c_T^{-3} \right)/3 \right]^{-1/3} \simeq 4.13$, and $\omega_D = q_D c_D \simeq 16.19$ are the Debye wave vector, velocity and frequency, respectively, with $c_L \simeq 8.71$ and $c_T \simeq 3.65$ the longitudinal and transverse sound velocities. The Debye vDOS is $g_D(\omega) = 3\,\omega^2/\omega^3_D$.
\begin{figure}[t]
\centering
\includegraphics[width=0.49\textwidth]{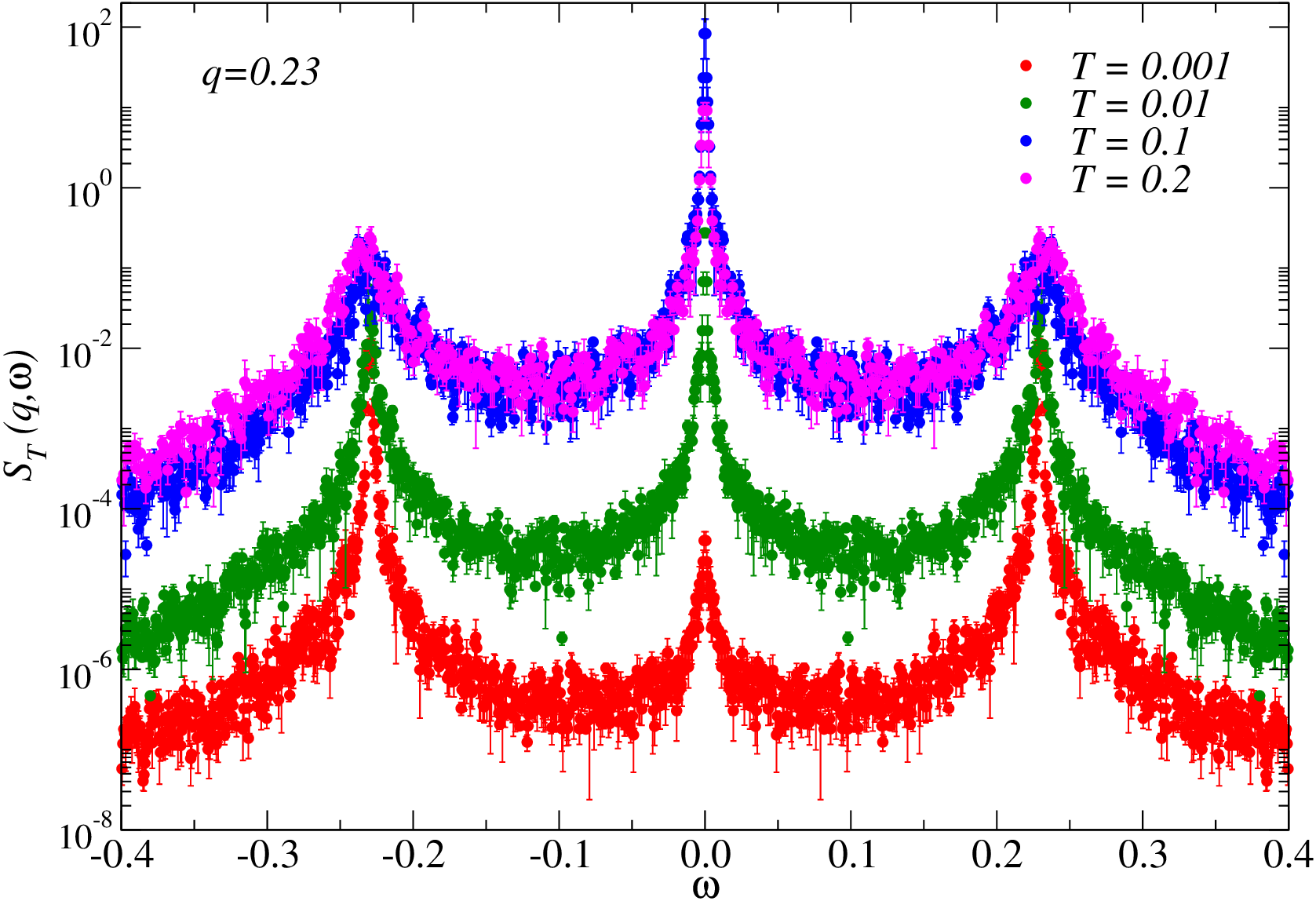}
\caption{
Transverse dynamical structure factors, $S_T(q,\omega)$, for the investigated LJ glass at the wave vector $q=0.23$. The different symbols correspond to the indicated temperatures, all below the glass-transition temperature, $T_g=0.4$.
}
\label{fig:spectra}
\end{figure}
%

\subsection{Calculation of local elastic constants}
In addition, we have characterized the degree of elastic heterogeneity as discussed in~\cite{Mizuno_2013}. It has been demonstrated that in LJ systems the bulk ($K$) and shear ($G$) moduli are such that $K\gg G$, and the latter mostly controls low-frequency transverse modes propagation~\cite{Mizuno2_2013,Mizuno_2014,Mizuno_2016}. We therefore focus on the probability distributions of $G$, determined by partitioning the box into an array of cubic domains of linear size $w \simeq 3.16$, identified by an index $m$ and including about $30$ particles each. The local moduli $G^m$ were computed by the fluctuation formula~\cite{yoshimoto_2004,lutsko_1988}, dubbed as the ``fully local" approach in~\cite{Mizuno_2013}. 

\section{Results}
The $S(q,\omega)$ spectra (see Fig.~\ref{fig:spectra} and Ref.~\cite{mizuno2019impact}) are characterized by two symmetric Brillouin peaks, flanking the elastic line. As $T$ increases, they move towards higher frequencies, with an increasing total intensity and broadening. We can extract quantitative information from these data by fitting the points in the spectral region around the Brillouin peaks to the damped harmonic oscillator model~\cite{sette1998dynamics}. This involves the parameters $\Omega(q)$ (related to the sound velocity by $c(q)=\Omega(q)/q$, see~Ref.~\cite{mizuno2019impact}), and $\Gamma(q)$, which encode the characteristic frequency and inverse life-time (or broadening, full width at half maximum) of the sound excitations, respectively.

\subsection{Sound damping}
\begin{figure}[t]
\centering
\includegraphics[width=0.49\textwidth]{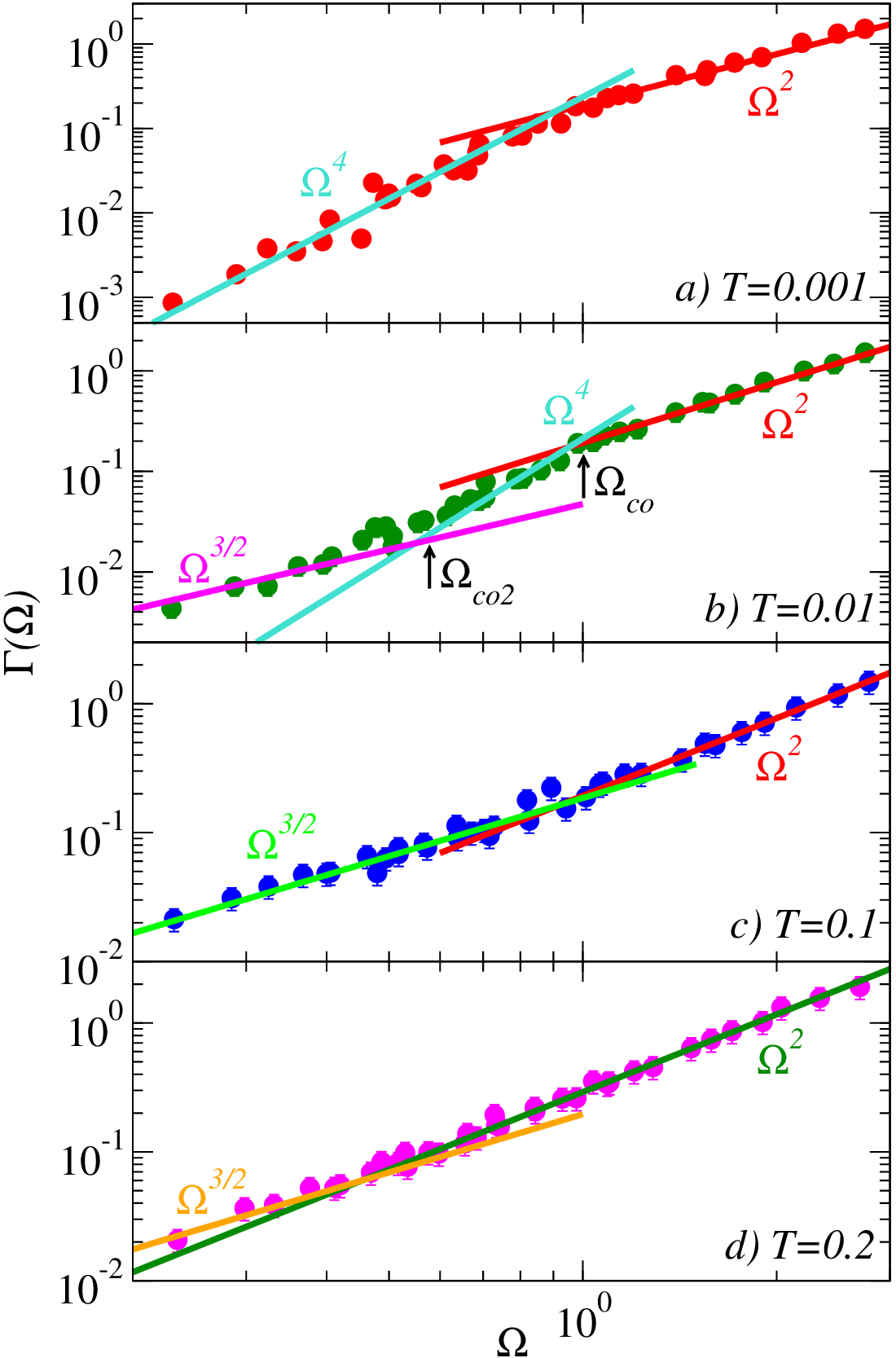}
\caption{
Transverse sound broadening, $\Gamma(\Omega)$ (circles), determined as discussed in the main text as a function of the corresponding Brillouin frequency, $\Omega$, at the indicated values of $T$. The solid lines indicate the power-law scaling valid in the different frequency regions, with  strongly $T$-dependent patterns as discussed in the text. We use the same color code to indicate power laws which arise from the same mechanism and are not modified at different values of $T$.
}
\label{fig:gamma}
\end{figure}

In Fig.~\ref{fig:gamma} we show the (total) $\Gamma$ as a function of the corresponding $\Omega$, at the indicated values of temperature. The $\Omega$-dependence of these data is very complex, and strongly depends on $T$. (We use symbols of the same color to identify the investigated temperatures, and solid lines of the same color for mechanisms that are not modified at different $T$-values.) At the lowest $T=10^{-3}$ (a), a clear crossover occurs between the high-frequency disorder-controlled behaviour $\propto \Omega^2$~\cite{ruocco1999nondynamic}, and a Rayleigh-like scattering contribution, $\propto\Omega^4$, at lower frequencies~\cite{angelani2000frustration}. As already noticed, the crossover frequency $\Omega_\text{co} \simeq 1$, is below the calculated Boson peak frequency, $\Omega_\text{BP}\simeq\Omega_\text{IR}\simeq 2$~\cite{monaco2009anomalous}. Note that even at this very low $T$ anharmonic interactions are obviously present and, for instance, still contribute to the thermal conductivity. Their intensity, however, is very low compared to other contributions, while non-negligible effects should be visible at frequencies smaller than our spectral range. By increasing $T$, in contrast, we expect the strength of anharmonicities to increase, eventually entering the frequency window.

This is indeed the case in (b) for $T=10^{-2}$, where we detect a second $T$-dependent crossover, at $\Omega_\text{co2} \simeq 0.6$, between the Rayleigh region and a remarkable low-frequency $\propto\Omega^{3/2}$ regime~\footnote{Note that a behaviour $\propto\Omega^{2}$ can also be adjusted to the same data. We have opted for the fractional exponent for consistence with the clearer observations at higher $T$.}, as theoretically predicted in Ref.~\cite{marruzzo2013vibrational} and reminiscent of the fractional attenuation of Refs.~\cite{ferrante2013acoustic,marruzzo2013vibrational} (see below). Note that the latter is obviously strongly $T$-dependent, whereas the Rayleigh and disorder-controlled regimes are not modified even at intermediate $T$, a feature that we will exploit below. Also, by increasing $T$, we expect the two crossover frequencies to eventually merge $\Omega_\text{co2}\simeq\Omega_\text{co}$, when the strength of the anharmonic couplings becomes comparable to that associated to the effect of the disorder, and the two mechanisms bury the Raleigh scaling in the entire $\Omega$-range. This is exactly what we observe at $T=10^{-1}$ in (c), where the $\propto \Omega^{3/2}$ regime at low frequency directly joins to the quadratic $T$-independent contribution at high $\Omega$.

\begin{figure}[t]
\centering
\includegraphics[width=0.49\textwidth]{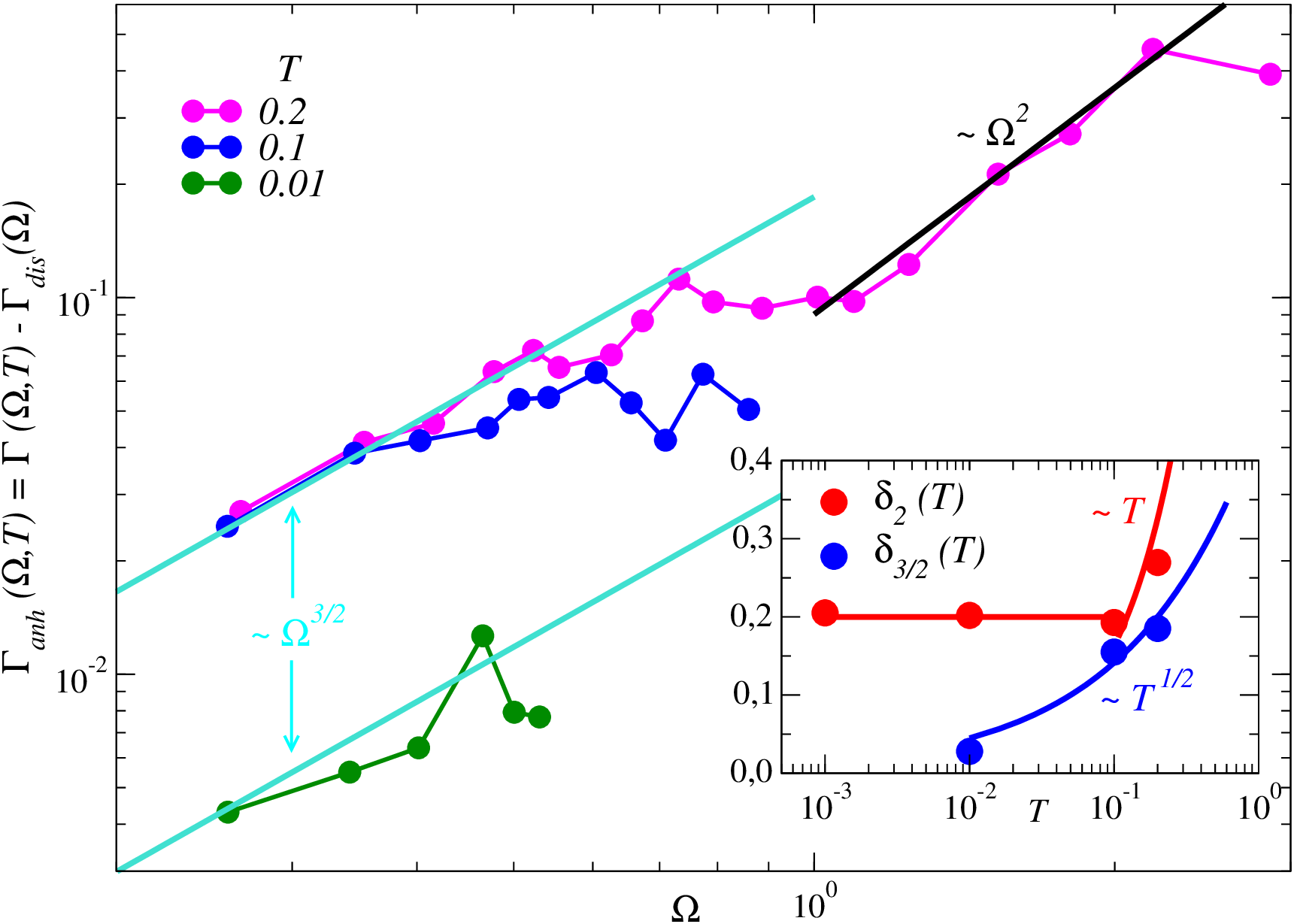}
\caption{
The anharmonic contribution $\Gamma_\text{anh}(\Omega,T)$ calculated by subtracting from the total broadening the disorder term at $T=10^{-3}$ related to the elastic heterogeneities. In the inset we show the $T$-dependence of the parameter $\delta_2$ and $\delta_{3/2}$, discussed in the main text.}
\label{fig:anharmonic}
\end{figure}

Finally, at the highest $T=2\times 10^{-1} \simeq T_g/2$ in (a), we observe a unique envelope of all scattering mechanisms which now scales uniformly as $\Omega^2$ (Akhiezer-like) in most of the $\Omega$-range, while a vestige of the $\Omega^{3/2}$ regime is still detected at low $\Omega$. It is worth to note that, at this stage, the width is fully $T$-dependent, and the second crossover $\Omega_\text{co2}$ shift towards lower frequency on increasing $T$, indicating that the Akhiezer-like $\Omega^2$-regime growths faster than the $\Omega^{3/2}$ region. Indeed, the pre-factor of the quadratic term $\delta_2(T)$, shown in the inset of Fig.~\ref{fig:anharmonic}, keeps a constant value for $T<0.1$, before substantially increasing (possibly linearly) at higher $T$.
 
The above scenario is similar to that reported recently in the experimental work of~\cite{Baldi_2014} for a network glass (sodium silicate), although in that case, a plain $\propto\Omega^2$ Akhiezer regime is reported instead of the fractional behaviour at low frequencies and high $T$. This extremely complex situation definitely points to non-trivial effects due to temperature on the sound waves propagation, which superimpose to $T$-independent effects of a completely different nature. In~Ref.~\cite{Mizuno_2014} we have related the latter to the existence of local elastic heterogeneities. To obtain additional quantitative insight we need at this point to disentangle the different contributions.
%

\subsection{Anharmonic contribution}
We start by posing $\Gamma(\Omega,T) = \Gamma_\text{dis}(\Omega) + \Gamma_\text{anh}(\Omega,T)$~\cite{ferrante2013acoustic}, where $\Gamma_\text{dis}(\Omega)$ encodes the $T$-independent effect of disorder and is related to the ($T=0$) inherent structure features. $\Gamma_\text{anh}(\Omega,T)$, in contrast, is a $T$-dependent contribution related to the anharmonic couplings at finite $T$. In all generality, we model $\Gamma_\text{dis}(\Omega)$ with a term $\propto\Omega^4$ in the Rayleigh, and $\propto\Omega^2$, in the high-frequency regions, respectively. We next join continuously the two power laws at $\Omega = \Omega_\text{co}$, by imposing 
\begin{equation} 
\label{eq:gammadis}
\Gamma_\text{dis}(\Omega) =
\left\{ \begin{aligned}
& \frac{\alpha_\text{dis}}{\Omega_\text{co}^{2}} \Omega^4, & \text{for}\ \ \Omega < \Omega_\text{co}, \\
& \alpha_\text{dis} \Omega^2, & \text{for} \ \ \Omega > \Omega_\text{co}.
\end{aligned} \right.
\end{equation}
By adjusting this formula to the data of Fig.~\ref{fig:gamma}~(a) we extract $\alpha_\text{dis} \simeq 0.2$ and $\Omega_\text{co} \simeq 1.0$. ($\Omega_\text{co}$ corresponds to the length scale $\xi=2\pi c_T/\Omega_\text{co}\simeq 20$, which is associated to the vortex-like structure of the non-affine displacement field~\cite{mizuno2019impact}.) {{\color{black}We can now obtain the anharmonic contribution by subtracting the disorder term from the total broadening, as
\begin{equation}
\Gamma_\text{anh}(\Omega,T) = \Gamma(\Omega,T) - \Gamma_\text{dis}(\Omega).
\end{equation}
Note that we have systematically smoothed the quite scattered data by averaging over bins each containing two points.

We show the results for $\Gamma_\text{anh}(\Omega,T)$ at $T\geq 10^{-2}$ in the main panel of Fig.~\ref{fig:anharmonic}. As expected, all curves vanish in the limit $\Omega\rightarrow 0$, where anharmonic effects must disappear. At $T=10^{-2}$ the fractional $\Omega^{3/2}$ dependence only survives at low frequencies, $\Omega \leq \Omega_\text{co2}$.} {\color{black}We observe an analogous behaviour at the higher $T=0.1$, although the intensity of the $\Omega^{3/2}$ term now increases of almost a factor of four. Eventually, at the highest $T=2\times 10^{-1}$, we also recover the $\Omega^{3/2}$ term which, however, crosses over to a residual $\Omega^2$ dependence for $\Omega \gtrsim \Omega_\text{co}$.} {\color{black} Note that the last feature is still of anharmonic origin, and is not related to a variation with temperature of the strength of the elastic heterogeneity, as already demonstrated by some of us in~\cite{schirmacher_2007}. Overall, the $\Omega^{3/2}$ scaling at low frequency confirms the fractional frequency dependence of broadening reported in the experimental work of~\cite{ferrante2013acoustic}, and predicted by the theory of~\cite{marruzzo2013vibrational}.

Finally we note that the complete ($T$-$\Omega$) dependence for the total $\Gamma$ in this low-frequency range has been proposed to scale as $\Gamma(\Omega,T)=\delta_{3/2}(T)\,\Omega^{3/2}$, with $\delta_{3/2}(T)\propto\sqrt{T}$~\cite{ferrante2013acoustic,marruzzo2013vibrational}, which also seems to be fulfilled by our data, as shown in the inset of Fig.~\ref{fig:anharmonic}. }

\subsection{Vibrational density of states}
We now demonstrate precisely and in a very direct way the relation between the anomalous transverse acoustic-like excitations behaviour and the BP properties. Note that in Ref.~\cite{monaco2009anomalous} we demonstrated a possible connection between the sound softening encoded in the pseudo-dispersion curves and the BP, simply assuming $q$ to be a good parameter for labelling vibrations in glasses, and counting the number of acoustic modes in the low-$q$ region. This procedure quite accurately reproduced the BP feature in the reduced $\hat{g}(\Omega)=g(\Omega)/\Omega^2$.

Here we adopt a different point of view, based on the heterogeneous elasticity theory~\cite{schirmacher_2006,schirmacher_2007,Marruzzo_2013,Schirmacher_2015}, which provides a remarkable relation between the (longitudinal) broadening and the BP of the form $\hat{g}(\Omega)=3/\Omega_D^3 +f\,\Gamma(\Omega)/(\Omega_D\Omega)^2$~\cite{schirmacher_2007}. In the original theory, $f$ ($\simeq 4.8$ in the present case) is a frequency-independent parameter that can be determined from the macroscopic velocities of sound and the density of the material. In the present work we consider a plainly $\Omega$ and $T$-dependent model, by considering the approximation of~\cite{Marruzzo_2013} for the response functions. One can show that, for $c_L > c_T$ and $\Gamma_L \ll \Gamma_T$, the analogous relation for the transverse broadening can be simply expressed as~\cite{mizuno2018phonon}
\begin{equation}
\hat{g}(\Omega) = \frac{3}{\Omega_D^3} + \left[ \frac{4}{\pi q_D^2 c^{2}(\Omega)} \right] \left[ \frac{\Gamma(\Omega)}{\Omega^2+\Gamma^2(\Omega)} \right].
\end{equation}
Note that with this model, we include both the effect of the sound softening, and a mild temperature dependence of the macroscopic velocity observed at high $T$~\cite{mizuno2019impact}.
\begin{figure}[t]
\centering
\includegraphics[width=0.49\textwidth]{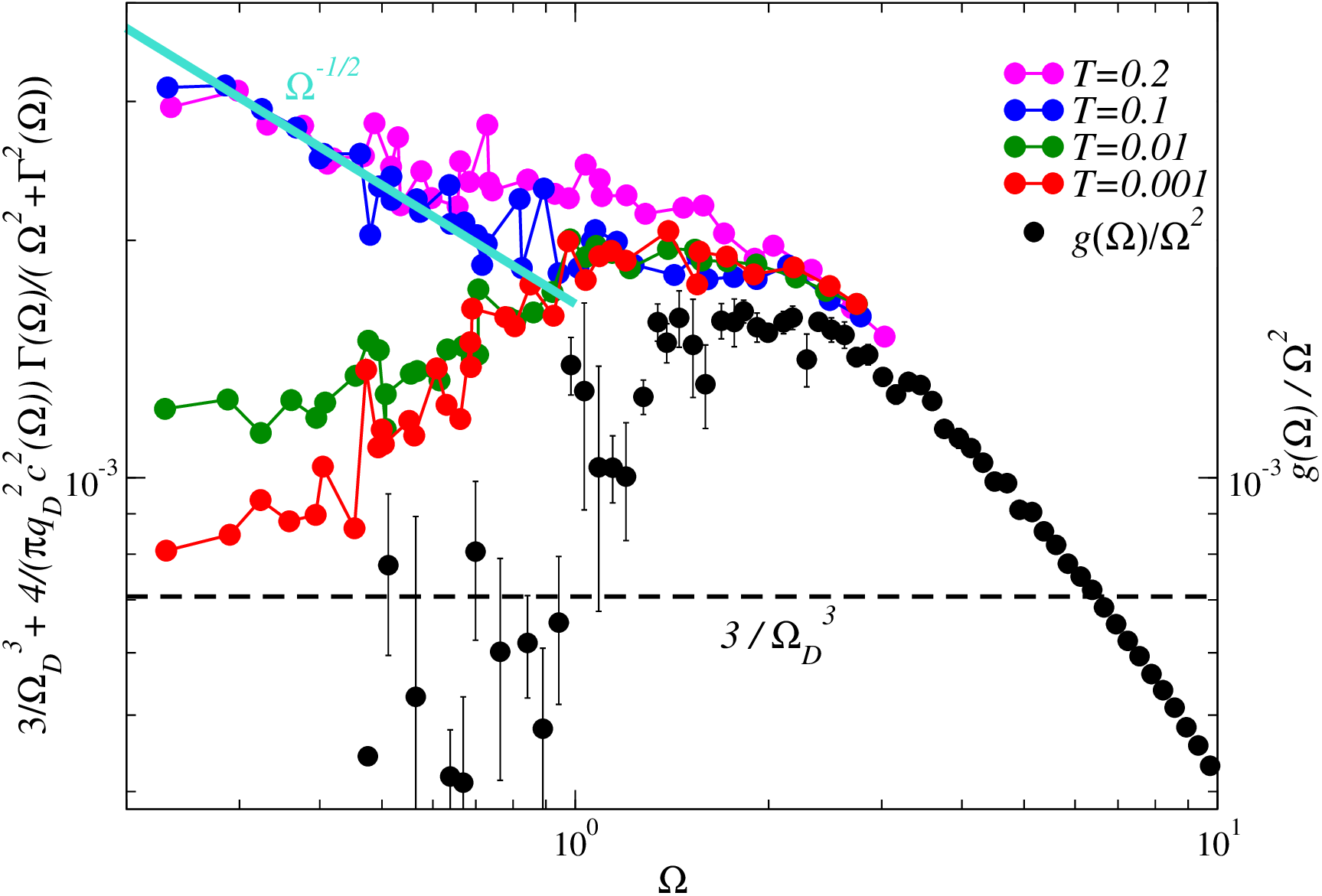}
\caption{
Comparison between the reduced density of state, $\hat{g}(\omega)$ (black) and the broadening data rescaled according to the heterogeneous elastic theory, at the indicated temperatures.
The horizontal dashed line shows the Debye limit, $3/\Omega_D^3$.
The solid line indicates the low-frequency fractional $\Omega^{-1/2}$ term related to anharmonicities. These data are discussed at length in the main text.}
\label{fig:theory}
\end{figure}

In Fig.~\ref{fig:theory} we plot separately the two sides of the equation discussed above, {\em without} any adjustable parameters. The data at $T=10^{-3}$ are in nice qualitative agreement with the $\hat{g}(\Omega)$. Indeed, the rescaled broadening data grasp the macroscopic (Debye) limit, increase quadratically in the Raleigh range, and saturate to a constant in the BP region, very close to the BP intensity. The situation is similar at $T=10^{-2}$, although anharmonicities already start to alter the small-$\Omega$ behaviour, a modification which is complete at the two highest temperatures. Now the data {\em decrease} by increasing frequency, following an $\Omega^{-1/2}$ power-law, eventually saturating at the BP.
We note that the consistent collapse of all data in this region is made possible by our more realistic model, which now also includes anharmonic modifications of the macroscopic velocities, as noticed above. Also, the simultaneous presence of both the $\Omega^{-1/2}$ dependence and the BP corroborates the predictions of~\cite{marruzzo2013vibrational}, where the theory was modified to include a small anharmonic scattering contribution~\cite{schirmacher2010sound,Tomaras_2010}, generating the same fractional behaviour of $\hat{g}(\omega)$. Similar data have been reported in the experimental work of~\cite{Baldi_2014}. 
\begin{figure}[t]
\centering
\includegraphics[width=0.49\textwidth]{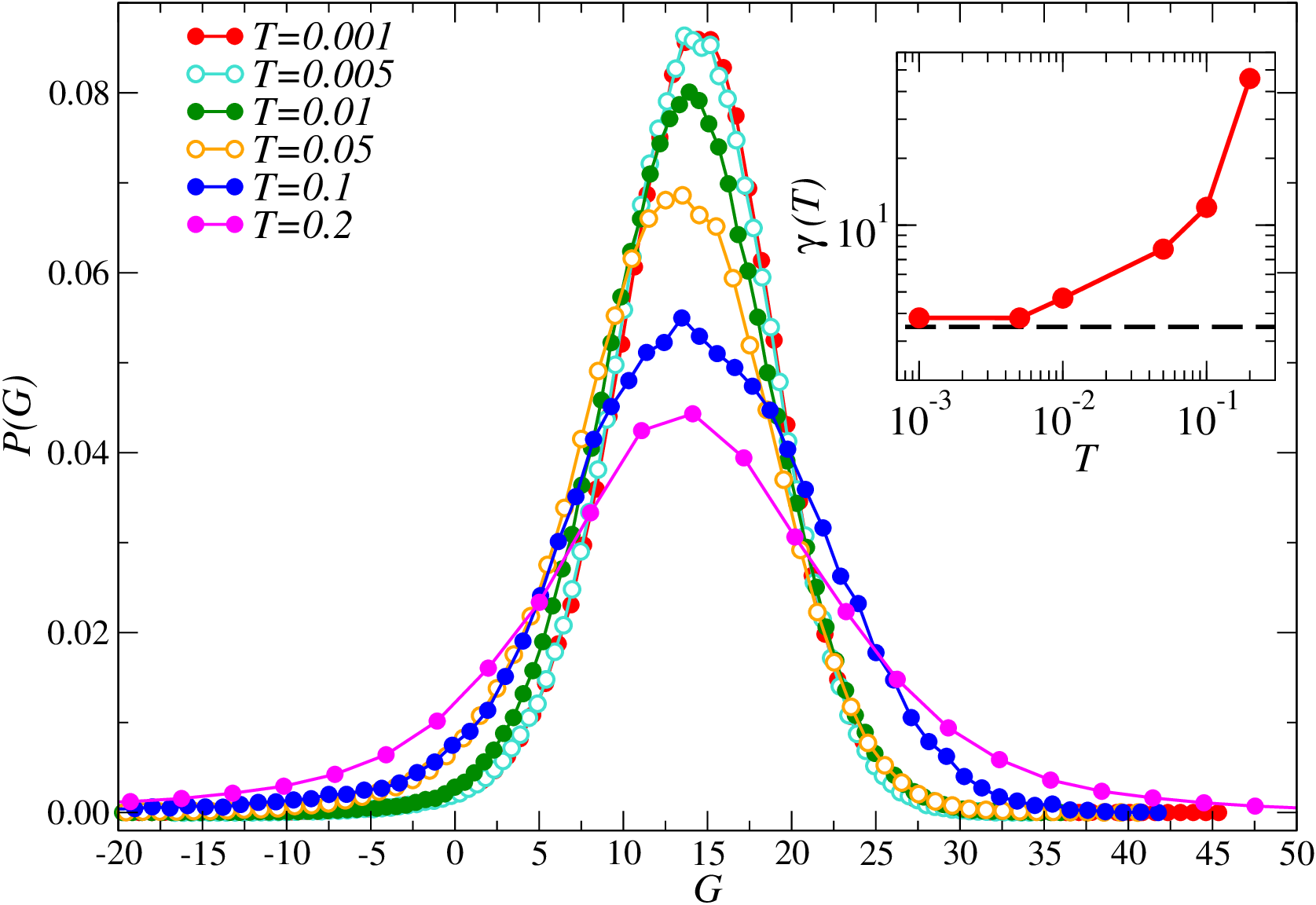}
\caption{
Probability distributions, $P(G)$, of the local shear moduli at the indicated values of $T$. The data corresponding to $T=0$ have been calculated in the harmonic limit~\cite{Mizuno2_2015}. All data, including the broadening with $T$ is discussed in the main text. In the inset we show (symbols) the $T$-dependence of the disorder parameter, $\gamma$, together with the calculated harmonic limit (dashed line). 
}
\label{fig:elastic}
\end{figure}
%

\subsection{Elastic heterogeneities}
We are now in the position to show directly that the described increase of the strength of the anharmonicities away from the harmonic limit is coupled to important modifications of the degree of the local elastic heterogeneity. Note that in the heterogeneous elasticity theory~\cite{schirmacher_2006,schirmacher_2007,Marruzzo_2013,Schirmacher_2015}, this feature is an {\em input}, which amounts to adjust the disorder parameter $\gamma=\hat{\rho} w^3\langle \delta G^2\rangle/\langle G\rangle^2$, related to the momenta of the shear moduli distributions, $P(G)$. In our simulations, in contrast, the heterogeneity can be {\em measured} by directly computing the $P(G)$, as recalled above.
We can therefore immediately confirm (see main panel in Fig.~\ref{fig:elastic}) the assumption that the local shear modulus is space-dependent, with Gaussian probability distributions, $P(G)$. Fluctuations in the local bulk modulus are, in contrast, negligible~\cite{Mizuno2_2013,Mizuno_2014,Mizuno_2016}.

As in Fig.~\ref{fig:elastic}, for $T<10^{-1}$, the $P(G)$ are very mildly $T$-dependent, with means and variances almost constant and very close to the $T=0$ harmonic values~\cite{mizuno2019impact}. At $T \ge 10^{-1}$, however, thermal fluctuations set in, strongly modifying the distributions. These broaden and include an increasing fraction of negative shear stiffnesses on increasing $T$. In the inset we show the $T$-dependence of $\gamma$ (symbols), which stays very close to the harmonic value $\gamma(T=0) \simeq 3.5$~\cite{mizuno2019impact} (dashed line) for $T<10^{-2}$. At higher temperatures, it starts to increase significantly, with a clear correlation with the Akhiezer-like linear increase of the strength $\delta_2(T)$ in the inset of Fig.~\ref{fig:anharmonic}. This behaviour therefore signals the approach to the elastic instability at the $\gamma_c$ of~\cite{marruzzo2013vibrational,ferrante2013acoustic}, with $\delta\gamma=\gamma-\gamma_c$ ($\le 0$) increasing with $T$ towards $0$. 

\section{Conclusion}
In this work we have elucidated the simultaneous impact of the anharmonic couplings and the effects due to disorder on the transverse sound waves propagation in glasses, with relative strengths determined by the temperature. Based on numerical data of unprecedented accuracy, we have provided a complete characterization of sound broadening, $\Gamma(\Omega,T)$, analyzing in depth the evolution of different scattering mechanisms in a very large frequency range. On one side, we have completely characterized the anharmonic channel, identifying a fractional frequency scaling predicted by the heterogeneous elasticity theory, modified to include anharmonic damping. On the other, we have convincingly linked the elastic moduli heterogeneities, which can be precisely quantified by simulation, to the cross-over from the Raleigh $\Omega^4$ to the $\Omega^2$ regimes, both of which are determined by disorder. 

We conclude with an observation. With a different theoretical point of view and based on an elastic network model, it has been proposed~\cite{Wyart_2010,DeGiuli_2014} that the weak connectivity of the particles (isostatic feature), due to the vicinity at the jamming transition point, induce non-affine effects which strongly impact the vibrational excitations. Although the origin of these effects is different from that assumed in the elastic heterogeneity theory, both frameworks share the view that features of the non-affine displacement field alter vibrational excitations in disordered solids. It is clear that only developments able to precisely integrate these mechanical aspects and a full treatment of the anharmonic couplings will be able to provide the complete picture for sound waves propagation in disordered solids.

\section*{Acknowledgments}
We thank G.~Monaco and W.~Schirmacher for feedback in the early stage of this work, and A.~Ikeda and M.~Shimada for useful discussions. H.~M. is supported by JSPS KAKENHI Grant Number 19K14670 and the Asahi Glass Foundation. S.~M. is supported by ANR-18-CE30-0019 (HEATFLOW).

%

%
\end{document}